\documentclass[11pt,twoside]{article}
\usepackage{atmp}

\copyrightnotice{2000}{4}{1231}{1257} 

\newtheorem{Def}{Def.}[section]
\newtheorem{Thm}[Def]{Theorem}
\newtheorem{Prp}[Def]{Proposition}
\newtheorem{Lemma}[Def]{Lemma}

\newtheorem{Corollary}[Def]{Corollary}
\newcommand{\Proof}{{\noindent \em{Proof. }}}
\newcommand{\QED}{\ \hfill $\FBox$ \\[1em]}

\newcommand{\spc}{\;\;\;\;\;\;\;\;\;\;}
\newcommand{\bra}{\mbox{$< \!\!$ \nolinebreak}}
\newcommand{\ket}{\mbox{\nolinebreak $>$}}

\newcommand{\FBox}{\rule{2mm}{2.25mm}}

\newcommand{\Z}{{\rm Z\kern-.35em Z}}
\newcommand{\bP}{{\rm I\kern-.15em P}}
\newcommand{\Q}{\kern.3em\rule{.07em}{.65em}\kern-.3em{\rm Q}}
\newcommand{\R}{{\rm I\kern-.15em R}}
\newcommand{\h}{{\rm I\kern-.15em H}}
\newcommand{\C}{\kern.3em\rule{.07em}{.55em}\kern-.3em{\rm C}}
\newcommand{\T}{{\rm T\kern-.35em T}}

\newcommand{\AI}{${\bf{A_{I}}}$}
\newcommand{\AII}{${\bf{A_{II}}}$}
\newcommand{\sgn}{{\mbox{sgn }}}

\begin{document}

\title{\Large{Absence of Static, Spherically Symmetric Black Hole Solutions for
Einstein-Dirac-Yang/Mills Equations with Complete Fermion Shells}}

\url{gr-qc/0005028}             

\author{Felix Finster}          
\address{Max Planck Institute for Mathematics in the Sciences \\
Leipzig, Germany}
\addressemail{Felix.Finster@mis.mpg.de}

\author{Joel Smoller}
\address{Mathematics Department \\
The University of Michigan, Ann Arbor, MI}
\addressemail{smoller@umich.edu}

\author{Shing-Tung Yau}
\address{Mathematics Department \\
Harvard University, Cambridge, MA}
\addressemail{yau@math.harvard.edu}

\markboth{\it ABSENCE OF BLACK HOLE SOLUTIONS}
{\it F.\ FINSTER, J.\ SMOLLER, S.-T.\ YAU}

\begin{abstract}
We study a static, spherically symmetric system of $(2j+1)$ massive
Dirac particles, each
    having angular momentum $j$, $j=1,2,\ldots$, in a classical gravitational
    and $SU(2)$ Yang-Mills field.  We show that for any black hole solution of
    the associated Einstein-Dirac-Yang/Mills equations, the spinors must vanish
    identically outside of the event horizon.
\end{abstract}

\cutpage

\section{Introduction}
\label{sec.1}
Recently the Einstein-Dirac-Yang/Mills (EDYM) equations
were studied for a static, spherically symmetric system of a Dirac particle interacting
with both a gravitational field and an $SU(2)$ Yang-Mills field~\cite{FSY1, FSY2}.
In these papers, the Dirac particle had no angular momentum, and we could make a
consistent ansatz for the Dirac wave function involving two real spinor
functions. In the present paper, we allow the Dirac particles to have non-zero
angular momentum $j$, $j=1,2,\ldots$. Similar to~\cite{FSY3}, we can build up
a spherically symmetric system out of $(2j+1)$ such Dirac particles.
In this case however, a reduction to real 2-spinors is no longer
possible, but we can obtain a consistent ansatz involving four real spinor
functions.

We show that the only black hole solutions of our 4-spinor EDYM equations are
those for which the spinors vanish identically outside the black hole; thus
these EDYM equations admit only the Bartnik-McKinnon (BM) black hole solutions
of the $SU(2)$ Einstein-Yang/Mills equations \cite{KM, SWY}.  This result
extends our work in \cite{FSY2} to the case with angular momentum; it again
means physically that the Dirac particles must either enter the black hole or
escape to infinity.  This generalization comes as a surprise because if one
thinks of the classical limit, then classical point particles with angular
momentum can ``rotate around'' the black hole on a stable orbit.  Our result
thus shows that the non-existence of black hole solutions is actually a quantum
mechanical effect.  A simple way of understanding the difference between
classical and quantum mechanical particles is that for classical particles, the
centrifugal barrier prevents the particles from falling into the black hole,
whereas quantum mechanical particles can tunnel through this barrier.  In our
system, tunnelling alone does not explain the non-existence of black-hole
solutions, because the Dirac particles are coupled to the classical fields; that
is, they can influence the potential barrier.  Our results are established by
analyzing in detail the interaction between the matter fields and the
gravitational field.

In Section~\ref{sec.2}, we derive the static, spherically symmetric $SU(2)$ 
EDYM equations with non-zero angular momentum of the Dirac particles.
By assuming the BM ansatz for
the YM potential (the vanishing of the electric component),
the resulting system consists of 4 first-order equations for the spinors, two 
first-order Einstein equations, and a second-order equation for the YM
potential. 
This EDYM system is much more complicated than the system considered 
in~\cite{FSY2}, and in order to make possible a rigorous
mathematical analysis of the 
equations, we often assume (as in~\cite{FSY3}) a power ansatz for 
the metric functions and the YM potential. Our analysis combines both
geometrical and analytic techniques.

\section{Derivation of the EDYM Equations} 
\label{sec.2}
We begin with the separation of variables for the 
Dirac equation in a static, spherically symmetric EYM background. As in
\cite{FSY1}, we choose the line element and the YM potential ${\cal{A}}$
in the form
\begin{eqnarray}
ds^2 &=& \frac{1}{T(r)^2} \:dt^2 \:-\: \frac{1}{A(r)} \:dr^2 \:-\:
r^2 \:d\vartheta^2 \:-\: r^2\: \sin^2 \vartheta \:d \varphi^2
\label{eq:2met} \\
{\cal{A}} &=& w(r) \:\tau^1\: d\vartheta \:+\: (\cos \vartheta \:
\tau^3 \:+\: w(r) \:\sin \vartheta \:\tau^2) \:d\varphi \label{eq:2pot}
\end{eqnarray}
with two metric functions $A$, $T$, and the YM potential $w$.
The Dirac operator was computed in \cite[Section~2]{FSY1} to be
\begin{eqnarray}
G &=& i T \:\gamma^t \partial_t \:+\: \gamma^r \left(i \sqrt{A} 
\partial_r + \frac{i}{r} \:(\sqrt{A}-1) -\frac{i}{2} \:\sqrt{A} 
\:\frac{T^\prime}{T} \right) \nonumber \\
&&\:+\: i \gamma^\vartheta 
\partial_\vartheta \:+\: i \gamma^\varphi \partial_\varphi
\:+\: \frac{2i}{r} \:(w-1) \:(\vec{\gamma} \vec{\tau}
- \gamma^r \tau^r) \:\tau^r
\;.\;\;\;\;\: \label{eq:1e}
\end{eqnarray}
This Dirac operator acts on 8-component wave
functions, which as in~\cite{FSY1} we denote by
$\Psi = (\Psi^{\alpha u a})_{\alpha, u, a=1,2}$, where $\alpha$ are the two spin
orientations, $u$ corresponds
to the upper and lower components of the Dirac spinor (usually called the
``large'' and ``small'' components, respectively), and $a$ is the YM index.
The Dirac equation is
\begin{equation}
    (G-m) \:\Psi \;=\; 0 \;, \label{eq:2dirac}
\end{equation}
where $m$ is the rest mass of the Dirac particle, which we assume to be
positive ($m>0$).

As explained in \cite{FSY1}, the Dirac operator (\ref{eq:1e})
commutes with the ``total angular momentum operators''
\begin{equation}
\vec{J} \;=\; \vec{L} \:+\: \vec{S} \:+\: \vec{\tau} \;,
    \label{eq:na}
\end{equation}
where $\vec{L}$ is angular momentum, $\vec{S}$ the spin 
operator, and $\vec{\tau}$ the standard basis of
${\mbox{su}}(2)_{\mbox{\scriptsize{YM}}}$. 
Thus the Dirac operator is invariant on the eigenspaces of total 
angular momentum, and we can separate out the angular dependence by 
restricting the Dirac operator to suitable eigenspaces of the 
operators $\vec{J}$. Since (\ref{eq:na}) can be regarded as the 
addition of angular momentum and two spins $\frac{1}{2}$, the 
eigenvalues of $\vec{J}$ are integers. In \cite{FSY1}, the Dirac equation 
was considered on the kernel of the operator $J^2$; this leads to the 
two-component Dirac equation \cite[(2.23),(2.24)]{FSY1}. Here we want to study 
the effect of angular momentum and shall thus concentrate on
the eigenspaces of $J^2$ with eigenvalues $j(j+1)$, $j=1,2,\ldots$.
Since the eigenvalues of $J_z$ merely describe the 
orientation of the wave function in space, it is furthermore 
sufficient to restrict attention to the eigenspace of $J_z$ corresponding to 
the highest possible eigenvalue. Thus we shall consider
the Dirac equation on the wave functions $\Psi$ with
\begin{equation}
J^2 \:\Psi \;=\; j(j+1)\:\Psi \spc {\mbox{and}} \spc
J_z \:\Psi \;=\; j \:\Psi \spc (j=1,2,\ldots).
    \label{eq:nb}
\end{equation}

Since (\ref{eq:nb}) involves only angular operators, it is convenient 
to analyze these equations on spinors $\Phi^{\alpha a}(\vartheta, 
\varphi)$ on $S^2$ (where $\alpha$ and $a$ are again the spin and YM indices, 
respectively). Let us first determine the dimension of the space spanned
by the vectors satisfying (\ref{eq:nb}). Using the well-known 
decomposition of two spins $\frac{1}{2}$ into a singlet and a triplet, 
we choose a spinor basis $\Phi_{st}$ with $s=0,1$ and $-s \leq t 
\leq s$ satisfying
\[ (\vec{S} + \vec{\tau})^2 \:\Phi_{st} \;=\; s(s+1) \:\Phi_{st} \;,\spc
(S_z + \tau_z) \:\Phi_{st} \;=\; t \:\Phi_{st} \;. \]
The spherical harmonics $(Y_{lk})_{l \geq 0, \:-l \leq k \leq l}$, on 
the other hand, are a basis of $L^2(S^2)$. Using the rules for the 
addition of angular momentum\cite{LL}, the wave functions 
satisfying~(\ref{eq:nb}) must be linear combinations of the 
following vectors,
\begin{eqnarray}
&& Y_{j\:j} \:\Phi_{0\:0} \label{eq:nc} \\
&& Y_{j-1 \:j-1} \Phi_{1\:1} \label{eq:nd} \\
&& Y_{j \:j-1} \Phi_{1\:1} \;\;\;,\spc Y_{j\:j} \Phi_{10} \label{eq:ne} \\
&& Y_{j+1 \:j-1} \Phi_{11} \;,\spc Y_{j+1 \:j} \Phi_{10}
\;,\spc Y_{j+1 \:j+1} \Phi_{1 \:-1} \;. \label{eq:nf}
\end{eqnarray}
These vectors all satisfy the second
equation in (\ref{eq:nb}), but they are not 
necessarily eigenfunctions of $J^2$. We now use the fact that a 
vector $\Psi \neq 0$ satisfying the equation $J_z \Psi = j \Psi$ is 
an eigenstate of $J^2$ with eigenvalue $j(j+1)$ if and only if it is 
in the kernel of the operator $J^+ = J_x + i J_y$.
Thus the dimension of the eigenspace (\ref{eq:nb}) coincides with the 
dimension of the kernel of $J^+$, restricted to the space spanned by the
vectors (\ref{eq:nc})-(\ref{eq:nf}). A simple calculation shows that this
dimension is four (for example, we have $J^+ \:(Y_{j \:j-1} \:\Phi_{1\:1}) =
Y_{j\:j} \:\Phi_{1 \:1} = J^+ \:(Y_{j \:j} \:\Phi_{1\:0})$, and thus 
$J^+$ applied to the vectors (\ref{eq:ne}) has a one-dimensional kernel).

We next construct a convenient basis for the angular functions 
satisfying (\ref{eq:nb}). We denote the vector (\ref{eq:nc}) by 
$\Phi_0$. It is uniquely characterized by the conditions
\begin{eqnarray*}
L^2 \:\Phi_0 & = & j(j+1) \:\Phi_0 \;,\spc\;\; L_z \:\Phi_0 \;=\; j \Phi_0 \\
(\vec{S}+\vec{\tau}) \:\Phi_0 & = & 0 \spc\;\;\;\;\;\;\:,\spc \|\Phi_0\|_{S^2} 
\;=\; 1 \;.
\end{eqnarray*}
We form the remaining three basis vectors by multiplying $\Phi_0$ with 
spherically symmetric combinations of the 
spin and angular momentum operators, namely
\begin{eqnarray*}
\Phi_1 &=& 2 S^r \:\Phi_0 \;=\; -2 \tau^r \:\Phi_0 \\
\Phi_2 &=& \frac{2}{c} \:(\vec{S} \vec{L}) \:\Phi_0 \;=\; -\frac{2}{c}
\:(\vec{\tau} \vec{L})\: \Phi_0 \\
\Phi_3 &=& \frac{4}{c}\:S^r \:(\vec{S} \vec{L}) \:\Phi_0
\;=\; -\frac{4}{c}\:\tau^r \:(\vec{\tau} \vec{L}) \:\Phi_0
\end{eqnarray*}
where
\[ c \;=\; \sqrt{j(j+1)} \;\neq\; 0 \;. \]
Since the operators $S^r$, $\tau^r$, and $(\vec{S} \vec{L})$ commute 
with $\vec{J}$, it is clear that the vectors $\Phi_1,\ldots,\Phi_3$ 
satisfy (\ref{eq:nb}). Furthermore, using the standard commutation relations
between the operators  $\vec{L}$, $\vec{x}$, and $\vec{S}$ \cite{LL},
we obtain the relations
\begin{eqnarray*}
(\vec{S} \vec{L})^2 &=& \frac{1}{4} \:L^2 \:+\: \frac{i}{2} \:
\epsilon_{jkl} \:S_l \:L_j L_k
\;=\; \frac{1}{4} \:L^2 \:-\: \frac{1}{4} \:
\epsilon_{jkl} \:S_l \:\epsilon_{jkm} L_m \\
&=& \frac{1}{4} \:L^2 \:-\: \frac{1}{2} \:\vec{S} \vec{L} \\
(\vec{S} \vec{\tau})^2 &=&
\frac{1}{4} \:\tau^2 \:-\: \frac{1}{2} \:\vec{S} \vec{\tau}
\;=\; \frac{3}{16} \:-\: \frac{1}{2} \:\vec{S} \vec{\tau} \\
2 S^r \:\Phi_0 &=& -2 \tau^r \:\Phi_0 \;=\; \Phi_1 \\
2 S^r \:\Phi_1 &=& -2 \tau^r \:\Phi_1 \;=\; \Phi_0 \\
2 S^r \:\Phi_2 &=& 2 \tau^r \:\Phi_2 \;=\; \Phi_3 \\
2 S^r \:\Phi_3 &=& 2 \tau^r \:\Phi_3 \;=\; \Phi_2 \\
(\vec{S} \vec{\tau}) \:\Phi_0 &=& -S^2 \:\Phi_0 \;=\; -\frac{3}{4} 
\:\Phi_0 \\
(\vec{S} \vec{\tau}) \:\Phi_1 &=& 2 S_k \tau_k \:S^r \:\Phi_0 \;=\;
-2 S_k \:(\vec{x} \vec{S})\:S_k \:\Phi_0 \\
&=& 2 S^2 \:S^r \:\Phi_0 - S_k x_k \:\Phi_0 \;=\; \frac{1}{2} \:S^r 
\:\Phi_0 \;=\; \frac{1}{4} \:\Phi_1 \\
(\vec{S} \vec{\tau}) \:\Phi_2 &=& \frac{2}{c} \:(\vec{S} \vec{\tau})
(\vec{S} \vec{L})\:\Phi_0 \;=\; \frac{1}{c} \:(\vec{L} \vec{\tau}) \:\Phi_0
- \frac{2}{c} \:(\vec{S} \vec{L}) (\vec{S} \vec{\tau}) \:\Phi_0 \\
&=& -\frac{1}{c} \:(\vec{S} \vec{L}) \:\Phi_0 \:+\: \frac{3}{2c} \:
(\vec{S} \vec{L}) \:\Phi_0 \;=\; \frac{1}{4} \:\Phi_2 \\
(\vec{S} \vec{\tau}) \:\Phi_3 &=& \frac{4}{c} \:(\vec{S} \vec{\tau}) \:S^r
(\vec{S} \vec{\tau})\:\Phi_0 \;=\; \frac{2}{c} \:\tau^r \:(\vec{S} \vec{L}) 
\:\Phi_0 - \frac{4}{c} \:S^r \:(\vec{S} \vec{\tau})^2 \:\Phi_0 \\
&=& \frac{1}{2} \:\Phi_3 \:-\: \frac{1}{2} \: S^r \:\Phi_2 \;=\; 
\frac{1}{4} \:\Phi_3 \\
(\vec{S} \vec{L}) \:\Phi_0 &=& \frac{c}{2} \:\Phi_2 \\
(\vec{S} \vec{L}) \:\Phi_1 &=& 2 \:(\vec{S} \vec{L}) \:S^r\:\Phi_0 \\
&=& 2 \:S_j \:[L_j, S^r] \:\Phi_0 \:+\: 2 \:\{S_j, S^r\} \:L_j 
\:\Phi_0 \:-\: 2 \:S^r\:(\vec{S} \vec{L}) \:\Phi_0 \\
&=& -2i\:S_j \:\epsilon_{jkl} \:x_k\:S_l\:\Phi_0 \:-\:\frac{c}{2} 
\:\Phi_3
\end{eqnarray*}
\begin{eqnarray*}
&=& -\epsilon_{jkl} \:x_j\:\epsilon_{klm} \:S_m\:\Phi_0 
\:-\:\frac{c}{2} 
\:\Phi_3 \;=\; -\Phi_1 \:-\:\frac{c}{2} \:\Phi_3 \\
(\vec{S} \vec{L}) \:\Phi_2 &=& \frac{2}{c} \:(\vec{S} \vec{L})^2 \:\Phi_0 \;=\;
\frac{2}{c} \left(\frac{1}{4} \:L^2 \:-\: \frac{1}{2} \:\vec{S} \vec{L}\right) 
\Phi_0 \;=\; \frac{c}{2} \:\Phi_0 \:-\:\frac{1}{2}\: \Phi_2 \\
(\vec{S} \vec{L}) \:\Phi_3 &=& \frac{4}{c} \:(\vec{S} \vec{L})\:S^r\:
(\vec{S} \vec{L}) \:\Phi_0
\;=\; -\frac{2}{c}\:S^r \:(\vec{S} \vec{L})\:\Phi_0 
\:-\:c\:S^r\:\Phi_0 \\
&=&-\frac{1}{2} \:\Phi_3 \:-\:\frac{c}{2}\:\Phi_1
\end{eqnarray*}
and thus
\begin{eqnarray*}
2 (\vec{S} \vec{\tau} - S^r \tau^r) \tau^r \:\Phi_0 &=& -\frac{1}{2} 
\:\Phi_1 \\
2 (\vec{S} \vec{\tau} - S^r \tau^r) \tau^r \:\Phi_1 &=& \frac{1}{2} 
\:\Phi_0 \\
2 (\vec{S} \vec{\tau} - S^r \tau^r) \tau^r \:\Phi_2 &=& 0 \\
2 (\vec{S} \vec{\tau} - S^r \tau^r) \tau^r \:\Phi_3 &=& 0 \;.
\end{eqnarray*}
Using these relations, it is easy to verify that the vectors 
$\Phi_0,\ldots,\Phi_3$ are orthonormal on $L^2(S^2)$.
We take for the wave function $\Psi$ the ansatz
\begin{eqnarray}
\lefteqn{ \Psi^{\alpha u a}(t,r,\vartheta,\varphi) } \nonumber \\
&=& e^{-i \omega t} \:\frac{\sqrt{T(r)}}{r} \left(
\alpha(r) \:\Phi_0^{\alpha a}(\vartheta, \varphi) \:\delta_{u,1}
\:+\:\gamma(r) \:\Phi_2^{\alpha a}(\vartheta, \varphi) \:\delta_{u,1} \right.
\nonumber \\
&& \left.\spc\;\;\;\;\;\;
\:+\: i \beta(r) \: \Phi_1^{\alpha a}(\vartheta, \varphi)
\:\delta_{u,2} \:+\: i \delta(r) \: \Phi_3^{\alpha a}(\vartheta, \varphi) 
\:\delta_{u,2} \right) \label{eq:Dans}
\end{eqnarray}
with real functions $\alpha$, $\beta$, $\gamma$, and $\delta$, where $\omega>0$
is the energy of the Dirac particle.  This is the simplest ansatz for which the
Dirac equation~(\ref{eq:2dirac}) reduces to a consistent set of ODEs.  Namely,
we obtain the following system of ODEs for the four-component
wave function $\Phi:=(\alpha, \beta, \gamma, \delta)$,
\begin{equation}
\sqrt{A} \:\Phi^\prime \;=\;
\left( \begin{array}{cccc}
\displaystyle \frac{w}{r} & -\omega T - m & \displaystyle \frac{c}{r} & 0 \\[.3em]
\omega T - m & \displaystyle -\frac{w}{r} & 0 & \displaystyle -\frac{c}{r}
\\[.3em]
\displaystyle \frac{c}{r} & 0 & 0 & -\omega T - m \\[.3em]
0 & \displaystyle -\frac{c}{r} & \omega T - m & 0
\end{array} \right) \Phi \;. \label{eq:2.1}
\end{equation}

Similar to~\cite{FSY3}, we consider the system of $(2j+1)$ Dirac wave functions
obtained from~(\ref{eq:Dans}) by applying the ladder operators $J_\pm$. 
Substituting the Einstein and YM equations~\cite{FSY1} and using the ansatz
(\ref{eq:2met}) and (\ref{eq:2pot}), we get the following system of ODEs,
\begin{eqnarray}
r A' &=& 1-A \:-\: \frac{1}{e^2}\: \frac{(1-w^2)^2}{r^2} \nonumber \\
&&\:-\: 2 \omega T^2 \:(\alpha^2+\beta^2+\gamma^2+\delta^2)
\:-\: \frac{2 A \:{w'}^2}{e^2} \label{eq:2.2}  \\
\frac{2 r A}{T}\: T' &=& A-1 \:+\: \frac{1}{e^2}\: \frac{(1-w^2)^2}{r^2} 
\nonumber \\
&&\:-\: 2 \omega  T^2 \:(\alpha^2+\beta^2+\gamma^2+\delta^2)
\:-\: \frac{2 A \:{w'}^2}{e^2} \nonumber \\
&&+ T \left[ 2m\:(\alpha^2-\beta^2+\gamma^2-\delta^2)
+ \frac{4c}{r}\:(\alpha \delta + \beta \gamma) \:+\: \frac{4w}{r} \:
\alpha \beta \right] \;\;\;\;\label{eq:2.3} \\
r^2 A\:w'' &=& -w(1-w^2) \:+\: e^2 \: r T \:\alpha \beta
\:-\: \frac{1}{2} \:r^2\: A'\: w' \:+\: \frac{r^2\: A \:T' \:w'}{T}
\;.
\label{eq:2.4}
\end{eqnarray}
Here (\ref{eq:2.2}) and (\ref{eq:2.3}) are the Einstein equations, and
(\ref{eq:2.4}) is the YM equation. Notice that the YM equation does not
depend on $\gamma$ and $\delta$; moreover the lower two rows in the Dirac
equation (\ref{eq:2.1}) are independent of $w$. This means that the Dirac
particles couple to the YM field only via the spinor functions $\alpha$
and $\beta$.
Indeed, a main difficulty here as compared to the two-spinor
problem~\cite{FSY2} will be to control the behavior of $\gamma$ and $\delta$.

For later use, we also give the equations for the following composite
functions,
\begin{eqnarray}
r^2 \:(A w')' &=& -w(1-w^2) \:+\: e^2\: r\: T\: \alpha \beta
\:+\: \frac{1}{2} \:r^2\: \frac{(A T^2)'}{T^2}\: w' \label{eq:2.5} \\
r \: (A T^2)' &=& \:-\:4 \omega \:T^4 \:(\alpha^2+\beta^2+\gamma^2+\delta^2)
\:-\: \frac{4 A \:T^2\:{w'}^2}{e^2} \nonumber \\
&&+  T^3 \left[ 2m\:(\alpha^2\!-\!\beta^2\!+\!\gamma^2\!-\!\delta^2)
+ \frac{4c}{r}\:(\alpha \delta + \beta \gamma) \:+\: \frac{4w}{r} \:
\alpha \beta \right] \!. \;\;\;\;\;\; \label{eq:2.6}
\end{eqnarray}
Also, it is quite remarkable and will be useful later that for $\omega=0$, the 
squared Dirac equation splits into separate equations for $(\alpha, \gamma)$ 
and $(\beta, \delta)$; namely from (\ref{eq:2.1}),
\begin{eqnarray}
\lefteqn{ \sqrt{A}\:\partial_r (\sqrt{A} \:\partial_r \Phi)  \;=\;
\left(m^2 + \frac{c^2}{r^2} \right) \Phi} \nonumber \\
&& \hspace*{-.5cm} +\left[ \sqrt{A} \:\left( \begin{array}{cccc}
\left(\frac{w}{r}\right)^\prime & 0 & -\frac{c}{r^2} & 0 \\
0 & -\left(\frac{w}{r}\right)^\prime & 0 & \frac{c}{r^2} \\
-\frac{c}{r^2} & 0 & 0 & 0 \\
0 & \frac{c}{r^2} & 0 & 0 \end{array} \right) +  \left( \begin{array}{cccc}
\frac{w^2}{r^2} & 0 & \frac{wc}{r^2} & 0 \\
0 & \frac{w^2}{r^2} & 0 & \frac{wc}{r^2} \\
\frac{wc}{r^2} & 0 & 0 & 0 \\
0 & \frac{wc}{r^2} & 0 & 0 \end{array} \right) \right] \Phi \;.
\;\;\;\;\;\;\;\label{eq:Ds}
\end{eqnarray}

\section{Non-Existence Results}
As in~\cite{FSY2}, we consider the situation where $r=\rho>0$
is the event horizon of a black hole, i.e.\ $A(\rho)=0$, and $A(\rho)>0$
if $r>\rho$. We again make (cf.\ \cite{FSY1})
suitable assumptions on the regularity of the event horizon:
\begin{description}
\item[(\AI)] The volume element $\sqrt{|\det g_{ij}}| = |\sin \vartheta| \:
r^2 A^{-1} \:T^{-2}$ is smooth and non-zero on the horizon; i.e.
\[ T^{-2} A^{-1}, \ T^2A \in C^1 ([\rho, \infty)) \;. \]
\item[(\AII)] The strength of the Yang-Mills field $F_{ij} $ is given by
\[ {\mbox{Tr}} (F_{ij} F^{ij}) \;=\; \frac{2A {w'}^2}{r^2}
+ \frac{(1-w^2)^2}{r^4} \;. \]
We assume that it is bounded near the horizon; i.e.
\begin{equation}
{\mbox{$w$ and $A {w'}^2$ are \ bounded for $\rho<r<\rho+\varepsilon$.}}
\label{2.AII}
\end{equation}
\end{description}
Furthermore, the spinors should be normalizable outside and away
from the event horizon, i.e.
\begin{equation}
\int_{\rho+\varepsilon}^{\infty} |\Phi|^2 \:\frac{T}{\sqrt{A}} \;<\; \infty
\spc{\mbox{for every $\varepsilon>0$}}.
    \label{eq:2.7}
\end{equation}
Finally, we assume that the metric functions and the YM potential satisfy a 
power ansatz near the event horizon. More precisely, setting
\[ u \;\equiv\; r-\rho \;, \]
we assume the ansatz
\begin{eqnarray}
A(r) &=& A_0 \:u^s \:+\: o(u^s) \label{eq:Ap} \\
w-w_0 &=& w_1 \:u^\kappa \:+\: o(u^\kappa) \label{eq:wp}
\end{eqnarray}
with real coefficients $A_0 \neq 0$ and $w_1$,
powers $s, \kappa>0$ and $w_0 = \lim_{r \searrow \rho} w(r)$.
Here and in what follows,
\[ f(u) \;=\; o(u^\nu) \;\;\;{\mbox{means that $\exists \delta>0$ with }}
\limsup_{r \searrow \rho} |u^{-\nu-\delta} \:f(u)| \;<\; \infty \;. \]
Also, we shall always assume that the derivatives of a function in $o(u^\nu)$ 
have the natural decay properties; more precisely,
\[ f(u) \;=\; o(u^\nu) \spc {\mbox{implies that}}\spc
f^{(n)}(u) \;=\; o(u^{\nu-n}) \;. \]
According to \AI, (\ref{eq:Ap}) yields that $T$ also satisfies a power law,
more precisely
\begin{equation}
T(r) \;\sim\; u^{-\frac{s}{2}} \:+\: o(u^{-\frac{s}{2}}) \;.
    \label{eq:Tp}
\end{equation}
Our main result is the following.
\begin{Thm} \label{thm:2.1}
Under the above assumptions, the only black hole solutions of the EDYM
equations (\ref{eq:2.1})--(\ref{eq:2.4}) are either the Bartnik-McKinnon
black hole solutions of the EYM equations, or
\begin{equation}
    s \;=\; \frac{4}{3} \spc{\mbox{and}}\spc \kappa \;=\; \frac{1}{3}\;.
    \label{eq:ex1}
\end{equation}
In (\ref{eq:ex1}), the so-called exceptional case, the spinors behave near the 
horizon like
\begin{equation}
(\alpha \beta)(r) \;\sim\; u^{\frac{1}{3}} + o(u^{\frac{1}{3}}) \;,\spc
0 \:<\: (\gamma \delta)(\rho) \:<\: \infty \;.
    \label{eq:ex2}
\end{equation}
\end{Thm}
Our method for the proof of this theorem is to assume a black hole solution
with $\Phi \not \equiv 0$, and to show that this implies (\ref{eq:ex1}) and 
(\ref{eq:ex2}). The proof, which is split up into several parts,
is given in Sections~\ref{sec4}--\ref{sec7}.

In Section~\ref{sec8}, we will analyze the exceptional case. It is shown 
numerically that the ansatz (\ref{eq:ex1}),(\ref{eq:ex2}) does not yield
global solutions of the EDYM equations. From this we conclude that for all
black hole solutions of our EDYM system, the Dirac spinors must vanish
identically outside of the event horizon.

\section{Proof that $\omega=0$}
\label{sec4}
Let us assume that there is a solution of the EDYM equations 
where the spinors are not identically zero, $\Phi \not \equiv 0$.
In this section we will show that then $\omega$ must be zero. 
First we shall prove that the norm of the spinors $|\Phi|$ is bounded from 
above and below near the event horizon. We distinguish between the two cases 
where $A^{-\frac{1}{2}}$ is or is not integrable near the event horizon.
\begin{Lemma} \label{lemma:3.1}
    If $A^{-\frac{1}{2}}$ is integrable near the event horizon $r=\rho$, then 
    there are positive constant $k$ and $\varepsilon$ such that
\begin{equation}
    \frac{1}{k}\;\leq\; |\Phi(r)|^2 \;\leq\; k \;,\spc
{\mbox{if $\rho<r<\rho+\varepsilon$.}}
    \label{eq:3.1}
\end{equation}
\end{Lemma}
{\Proof}
Writing (\ref{eq:2.1}) as $\sqrt{A}\: \Phi' = M \Phi$, we have
\begin{eqnarray}
\lefteqn{ \frac{1}{2}\: \sqrt{A} \: \frac{d}{dr} |\Phi|^2 \;=\;
    \frac{1}{2} \: \Phi^t (M+M^*) \Phi } \nonumber  \\
     & = & \frac{w}{r}\: (\alpha^2-\beta^2) \:-\: 2m \:(\alpha \beta + \gamma 
     \delta) \:+\: \frac{2c}{r}\:(\alpha \gamma - \beta \delta) \nonumber  \\
     & \leq & \frac{w}{r}\: (\alpha^2-\beta^2) \:+\:m \:(\alpha^2 + \beta^2 + 
     \gamma^2 + \delta^2) \:+\:
     \frac{2c}{r}\:(\alpha \gamma - \beta \delta) \nonumber  \\
     & \leq & c_1 \: |\Phi|^2 \;.
    \label{eq:3.2}
\end{eqnarray}
Here the constant $c_1$ is independent of $r \in (\rho, \rho+1]$,
since $w$ is bounded near the horizon according to assumption \AII.
Since we are assuming that $\Phi \not \equiv 0$ in $r>\rho$, the uniqueness
theorem for solutions of ODEs yields that $|\Phi|^2>0$ on $(\rho, 
\rho+1]$. Then dividing~(\ref{eq:3.2}) by $\frac{1}{2} \sqrt{A} |\Phi|^2$ and 
integrating from $r_1$ to $r_2$, $\rho<r_1<r_2$, we get
\[ \left|\log |\Phi(r_2)|^2 \:-\: \log |\Phi(r_1)|^2 \right| \;\leq\; 2 
c_1 \int_{r_1}^{r_2} A^{-\frac{1}{2}}(r)\:dr \;. \]
Taking the limit $r_1 \searrow \rho$ in this last inequality gives the 
desired result.
\QED

\begin{Lemma}
\label{lemma:3.15}
If $A^{-\frac{1}{2}}$ is not integrable near the event horizon $r=\rho$ and 
$\omega \neq 0$, then there are positive constants $k$ and $\varepsilon$
such that
\begin{equation}
\frac{1}{k} \;\leq\; |\Phi(r)|^2 \;\leq\; k \spc
{\mbox{if $\rho<r<\rho+\varepsilon$.}}
    \label{eq:3.69a}
\end{equation}
\end{Lemma}
{\Proof}
Define the matrix $J$ by
\[ J \;=\; \left( \begin{array}{cccc}
\displaystyle 1-\frac{m}{\omega T} & \displaystyle -\frac{w}{r \omega T} &
0 & \displaystyle -\frac{c}{r \omega T} \\[.9em]
\displaystyle -\frac{w}{r \omega T} & \displaystyle 1+\frac{m}{\omega T} &
\displaystyle -\frac{c}{r \omega T} & 0 \\[.9em]
0 & \displaystyle -\frac{c}{r \omega T} & \displaystyle 1-\frac{m}{\omega T} &
0 \\[.9em]
\displaystyle -\frac{c}{r \omega T} & 0 & 0 & \displaystyle 1+\frac{m}{\omega T}
\end{array} \right) \;\;\, \]
and notice that, since $T(r) \to \infty$ as $r \searrow \rho$,
$J$ is close to the identity matrix for $r$ near $\rho$. If we let
\[ F(r) \;=\; \bra \Phi(r),\:J(r)\:\Phi(r) \ket\;, \]
then a straightforward calculation yields that
\[ F' \;=\; \bra \Phi(r),\:J'(r)\:\Phi(r) \ket\;. \]
In a manner similar to that in~\cite{FSY2}, we can prove that $|J'|$ is 
integrable near $r=\rho$, and as in~\cite{FSY2}, it follows that~(\ref{eq:3.69a}) holds.
\QED

\begin{Lemma}
If $\Phi \not \equiv 0$ for $r>\rho$, then $\omega=0$.
\end{Lemma}
{\Proof}
Assume that $\omega \neq 0$. We write the $(AT^2)'$ equation (\ref{eq:2.6}) as
\begin{eqnarray}
\lefteqn{ r\: (A T^2)' \;=\; -4 \omega \:T^4 \:|\Phi|^2 \:-\: \frac{4 A 
{w'}^2}{e^2}\:T^2 } \nonumber \\
&& \:+\: \left[ 2 m\:(\alpha^2-\beta^2+\gamma^2-\delta^2) 
+\frac{4  \:w}{r} \:\alpha \beta
+\frac{4c}{r}\:(\alpha \delta + \beta \gamma) \right] T^3 \;. \label{eq:4y}
\end{eqnarray}
According to hypothesis \AII, the left side of this equation is bounded near
the event horizon. The Lemmas~\ref{lemma:3.1} and~\ref{lemma:3.15}
together with \AII\ imply that the coefficients of $T^4$, $T^3$, and $T^2$
in this equation are all all bounded, and that the coefficient of $T^4$
is bounded away from zero near $r=\rho$.
Assumption \AII\ implies that $T(r) \to \infty$ as $r \searrow 
\rho$. Hence the right side of (\ref{eq:4y}) diverges as
$r \searrow \rho$. This is a contradiction.
\QED

\section{Reduction to the Case $\alpha(\rho)=0$, $\beta(\rho) \neq 0$}
\label{sec5}
Since $\omega=0$, the Dirac equation (\ref{eq:2.1}) reduces to
\begin{equation}
    \sqrt{A}\: \Phi^\prime \;=\; \left( \begin{array}{cccc}
    w/r & -m & c/r & 0 \\ -m & -w/r & 0 & -c/r \\
    c/r & 0 & 0 & -m \\ 0 & -c/r & -m & 0 \end{array} \right) \Phi
\;\equiv\; M \Phi \;.
    \label{eq:3.3}
\end{equation}
The following Lemma gives some global information on the behavior of
the solutions to (\ref{eq:3.3}).
\begin{Lemma} \label{lem:3.4}
The function $(\alpha \beta + \gamma \delta)$ is strictly positive,
decreasing, and tends to zero as $r \to \infty$.
\end{Lemma}
{\Proof}
A straightforward calculation gives
\[ \sqrt{A} \:(\alpha \beta + \gamma \delta)' \;=\; -m \:|\Phi|^2 \;, \]
so that $(\alpha \beta + \gamma \delta)(r)$ is a strictly decreasing 
function, and thus has a (possibly infinite) limit as $r \to \infty$.
Since $|\Phi|^2 \geq 2 \:|\alpha \beta + \gamma \delta|$, we see that the
normalization condition (\ref{eq:2.7}) holds only if this limit is zero.
It follows that $(\alpha \beta + \gamma \delta)$ is strictly positive.
\QED

Next we want to show that the spinors have a (possibly infinite) limit as $r 
\searrow \rho$. When $A^{-\frac{1}{2}}$ is integrable near the event horizon,
it is an immediate consequence of Lemma~\ref{lemma:3.1} that this limit
exists and is even finite.
\begin{Corollary} \label{cor52}
If $A^{-\frac{1}{2}}$ is integrable near the horizon, then $\Phi$ has a
finite limit for $r \searrow \rho$.
\end{Corollary}
{\Proof}
We can integrate (\ref{eq:3.3}) from $r_1$ to $r_2$, $\rho<r_1<r_2$,
\[ \Phi(r_2) - \Phi(r_1) \;=\; \int_{r_1}^{r_2} A^{-\frac{1}{2}}(r) \: M(r) 
\:\Phi(r) \:dr \;. \]
Lemma~\ref{lemma:3.1} yields that the right side converges as $r_1 \searrow 
\rho$, and hence $\Phi$ has a finite limit.
\hspace*{4cm} \QED
In the case when $A^{-\frac{1}{2}}$ is not integrable near the horizon, we 
argue as follows. According to the power ansatz (\ref{eq:wp}), the matrix in
(\ref{eq:3.3}) has a finite limit on the horizon. Exactly as shown
in~\cite[Section~5]{FSY4} using 
the stable manifold theorem, there are fundamental solutions of the Dirac 
equation (that is, a basis of solutions of the ODE~(\ref{eq:3.3}))
which behave near the event horizon exponentially like 
$\exp(\lambda_j \int A^{-\frac{1}{2}})$, where $\lambda_j \in \R$ are the
eigenvalues for $r \searrow \rho$ of the matrix in (\ref{eq:3.3}) (notice
that the $\lambda_j$ are real since they are the eigenvalues of a symmetric matrix).
Thus for any linear 
combination of these fundamental solutions, the spinor functions are monotone 
in a neighborhood of the event horizon, and hence as $r \searrow \rho$,
$\Phi$ has a limit in $\R \cup \{\pm \infty \}$. We set
\[ \Phi(\rho) \;=\; \lim_{r \searrow \rho} \Phi(r) \;,\spc
(\alpha \beta)(\rho) \;=\; \lim_{r \searrow \rho} (\alpha \beta)(r) \;. \]
\begin{Prp} $(\alpha \beta)(\rho) =0$. \label{prp:5.3}
\end{Prp}
{\Proof}
We consider the $(A w')'$ equation (\ref{eq:2.5}) with $\omega=0$,
\begin{equation}
r^2 \:(A w')' \;=\; -w(1-w^2) \:+\: e^2\: r\: (\sqrt{A} T)\: \frac{\alpha 
\beta}{\sqrt{A}} \:+\: \frac{r^2 \:(A T^2)'}{2 A T^2} \:A w'\;.
    \label{eq:3.5}
\end{equation}
Suppose that
\begin{equation}
(\alpha \beta)(\rho) \;>\; 0 \;.
    \label{eq:3.6}
\end{equation}
From hypotheses \AI\ and and \AII, we se that the coefficient of $\alpha \beta
A^{-\frac{1}{2}}$ is positive near $r=\rho$, and the other terms on the right side
of~(\ref{eq:3.5}) are bounded. Thus we may write~(\ref{eq:3.5}) in the form
\begin{equation}
(A w')' \;=\; \Phi(r) \:+\: \frac{\Psi((r)}{\sqrt{A(r)}} \;,
    \label{eq:3.7}
\end{equation}
where $\Phi$ is bounded and $\Psi>0$ near $\rho$. Thus we can find constants 
$\Phi_0$, $\Psi_0$ satisfying
\begin{equation}
(A w')' \,>\, \Phi_0 \:+\: \frac{\Psi_0}{\sqrt{A(r)}} \;,\spc \Psi_0>0,
    \label{eq:3.8}
\end{equation}
for $r$ near $\rho$. Then exactly as in \cite[Section~3]{FSY2}, it follows
that the spinors must vanish in $r>\rho$.

If on the other hand
\begin{equation}
(\alpha \beta)(\rho) \;<\; 0 \;.
    \label{eq:3.9}
\end{equation}
then (\ref{eq:3.7}) holds with $\Psi(r)<0$ near $\rho$. Thus
\begin{equation}
-(A w')' \;=\; -\Phi(r) \:-\: \frac{\Psi((r)}{\sqrt{A(r)}} \;.
    \label{eq:3.10}
\end{equation}
Setting $\tilde{w}=-w$, (\ref{eq:3.10}) becomes
\[ (A \tilde{w}')' \;=\; -\Phi(r) \:-\: \frac{\Psi((r)}{\sqrt{A(r)}} \;, \]
where $-\Psi(r)>0$ for $r$ near $\rho$. Thus we see that (\ref{eq:3.7}) holds 
for $w$ replaced by $\tilde{w}$. This again leads to a contradiction.
\QED

The next proposition rules out the case that both $\alpha$ and $\beta$ vanish 
on the event horizon.
\begin{Prp} \label{prp53}
Either $\alpha(\rho)=0$, $\beta(\rho) \neq 0$ or
$\alpha(\rho) \neq 0$, $\beta(\rho)=0$.
\end{Prp}
{\Proof}
Suppose that
\begin{equation}
\alpha(\rho) \;=\; 0 \;=\; \beta(\rho) \;.
    \label{eq:3.17a}
\end{equation}
According to Lemma~\ref{lem:3.4}, $\gamma$ and $\delta$ cannot both vanish on
the event horizon.  Using~(\ref{eq:3.3}), we have for $r$ near $\rho$,
\begin{eqnarray}
    \sqrt{A}\: \alpha' & = & \frac{c}{r} \:\gamma \:+\: o(1)
    \label{eq:3.15a}  \\
    \sqrt{A}\: \beta' & = & -\frac{c}{r} \:\delta \:+\: o(1) \;.
    \label{eq:3.15b}
\end{eqnarray}
If $A^{-\frac{1}{2}}$ is not integrable near the event horizon, these equations
show that $\gamma(\rho)$ and $\delta(\rho)$ are finite (otherwise
multiplying (\ref{eq:3.15a}) and (\ref{eq:3.15b}) by $A^{-\frac{1}{2}}$ and
integrating would contradict (\ref{eq:3.17a})); if $A^{-\frac{1}{2}}$ is
integrable near $\rho$, Corollary~\ref{cor52} shows that $\gamma(\rho)$ and
$\delta(\rho)$ are again finite.

From (\ref{eq:2.6}) with $\omega=0$ we have
\begin{eqnarray}
r\: (A T^2)' &=& \left[ 2 m\:(\alpha^2-\beta^2+\gamma^2-\delta^2) 
+\frac{4  w}{r} \:\alpha \beta
+\frac{4c}{r}\:(\alpha \delta + \beta \gamma) \right] T^3 \nonumber \\
&& \:-\: \frac{4}{e^2} \: (A {w'}^2) \:T^2 \;.    \label{eq:3.12}
\end{eqnarray}
Since the coefficients of $T^3$ and $T^2$ are bounded, as is the 
left-hand side, we conclude that, since $T(r) \to \infty$ as $r 
\searrow \rho$, the coefficient of $T^3$ must vanish on the horizon,
\begin{equation}
    \left[ 2 m\:(\alpha^2-\beta^2+\gamma^2-\delta^2) 
    +\frac{4  w}{r} \:\alpha \beta
    +\frac{4c}{r}\:(\alpha \delta + \beta \gamma) \right] _{r=\rho} \;=\; 0 \;.
    \label{eq:4z}
\end{equation}
As a consequence, $\gamma(\rho)^2 = \delta(\rho)^2$, and
Lemma~\ref{lem:3.4} yields that
\begin{equation}
\gamma(\rho) \;=\; \delta(\rho) \;\neq\; 0 \;. \label{eq:3.16}
\end{equation}
Furthermore from (\ref{eq:3.15a}) and (\ref{eq:3.15b}), for $r$ near $\rho$,
\begin{equation}
\sgn \alpha(r) \;=\; \sgn \gamma(r) \;\;\;{\mbox{and}}\;\;\;
\sgn \beta(r) \;=\; -\sgn \delta(r)\;.
    \label{eq:3.17}
\end{equation}

\begin{figure}[tb]
        \centerline{\epsfbox{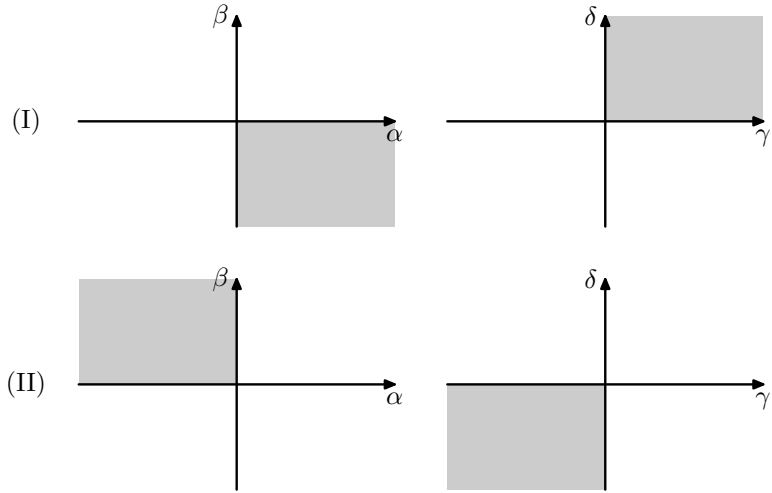}}
        \caption{Invariant Regions for the Spinors}
        \label{fig1}
\end{figure}

From (\ref{eq:3.16}) and (\ref{eq:3.17}), we see that for $r$ near $\rho$, 
the spinors must lie in the shaded areas in one of the two 
configurations (I) or (II) in Figure~\ref{fig1}.
Now we claim that in either configuration (I) or (II), the shaded regions
are invariant. For the proof, we consider the Dirac equation (\ref{eq:3.3}).
One easily checks that the shaded regions in the $\alpha/\beta$-plots
are invariant, provided that $\gamma$ and $\delta$ are as depicted in their 
shaded regions. Similarly, one verifies that the shaded regions in the
$\gamma/\delta$-plots are invariant, provided that $\alpha$ and $\beta$ lie
in the shaded regions. Moreover, Lemma~\ref{lem:3.4} shows that the spinors
cannot leave their regions simultaneously (i.e. for the same $r$).
This proves the claim.

Next we consider the situation for large $r$. In the limit $r \to 
\infty$, the matrix $M$ in (\ref{eq:3.3}) goes over to the matrix $S$ given by
\[ S \;=\; \left( \begin{array}{cccc}
    0 & -m & 0 & 0 \\ -m & 0 & 0 & 0 \\
    0 & 0 & 0 & -m \\ 0 & 0 & -m & 0 \end{array} \right) \;. \]
In $S$, the non-zero $2 \times 2$ upper and lower triangular blocks,
\[ \left( \begin{array}{cc}
    0 & -m \\ -m & 0 \end{array} \right) \;, \]
have eigenvectors $(1,1)^t$ and $(1,-1)^t$ with corresponding 
eigenvalues $-m$ and $m$, respectively. Since the system of ODEs
\[ \sqrt{A} \:\Phi' \;=\; S \:\Phi \]
splits into separate equations for
$(\alpha, \beta)$ and $(\gamma, \delta)$, we see that
$(\alpha(r), \beta(r))$ must be
a linear combination of $e^{-c(r) \:r} \:(1,1)^t$ and
$e^{d(r) \:r} \:(1,-1)^t$, where the functions
$c$ and $d$ are close to $m$. Since the 
spinors are assumed to be normalizable (i.e.\ (\ref{eq:2.7}) holds), and are 
non-zero for $r>\rho$, it follows that for large $r$, the spinors are close
to a constant multiple of $e^{-c(r) \:r} \:(1,1)^t$, and thus for large $r$,
$\sgn \alpha(r) = \sgn \beta(r)$. Similarly, for large $r$,
$\sgn \gamma(r) = \sgn \delta(r)$. This is a contradiction to the shaded
invariant regions of Figure~\ref{fig1}.
\hspace*{4cm} \QED

The two cases in Proposition~\ref{prp53} can be treated very similarly.
Therefore we shall in what follows restrict attention to the first case.
Furthermore, we know from Lemma~\ref{lem:3.4} and Proposition~\ref{prp:5.3}
that $(\gamma \delta)(\rho)>0$. 
Using linearity of the Dirac equation, we can assume that both
$\gamma(\rho)$ and $\delta(\rho)$ are positive. Hence the remaining
problem is to consider the case where
\begin{equation}
    \alpha(\rho)\;=\;0 \;,\spc \beta(\rho) \;\neq\; 0 \;,\spc
    \gamma(\rho),\:\delta(\rho) \;>\; 0 \;.
    \label{eq:5zz}
\end{equation}

\section{Proof that $A^{-\frac{1}{2}}$ is Integrable Near the Event Horizon}
\label{sec6}
In this section we shall assume that $A^{-\frac{1}{2}}$ is not integrable near 
the event horizon and deduce a contradiction. We work with the power ansatz 
(\ref{eq:Ap}),(\ref{eq:wp}) and thus assume that $s \geq 2$.

We first consider the case $w_0 \neq 0$. The first component of the squared 
Dirac equation (\ref{eq:Ds}) is
\begin{eqnarray}
\lefteqn{ \sqrt{A}\:\partial_r (\sqrt{A}\:\partial_r \alpha) } \nonumber \\
&=& \left[ m^2 \:+\: \frac{c^2 + w^2}{r^2} \:+\: \sqrt{A} 
\left(\frac{w}{r}\right)^\prime \right] \alpha \:+\: \frac{c 
\:(w-\sqrt{A})}{r^2}\:\gamma \;.
    \label{eq:6a}
\end{eqnarray}
The square bracket is bounded according to \AII. Since $\alpha(\rho)=0$ and 
$\gamma(\rho)>0$, our assumption $w_0 \neq 0$ implies that the right side of 
(\ref{eq:6a}) is bounded away from zero near the event horizon, i.e.\ there 
are constants $\delta, \varepsilon$ with
\[ \pm \sqrt{A} \:\partial_r (\sqrt{A} \:\partial_r \alpha) \;\geq\; \delta \spc
{\mbox{for $\rho < r < \rho+\varepsilon$}}, \]
where ``$\pm$'' corresponds to the two cases $w_0>1$ and $w_0<1$, respectively. 
We multiply this inequality by $A^{-\frac{1}{2}}$ and integrate from $r_1$ to 
$r_2$, $\rho<r_1<r_2$,
\[ \left. \pm \sqrt{A} \:\partial_r \alpha \right|_{r_1}^{r_2} \;\geq\; \delta \:
\int_{r_1}^{r_2} A^{-\frac{1}{2}} \;. \]
The right side diverges as $r_1 \searrow \rho$, and thus $\lim_{r \searrow 
\rho}\: \sqrt{A} \partial_r \alpha = \mp \infty$. Hence near the event horizon, 
$\mp \partial_r \alpha \geq A^{-\frac{1}{2}}$, and integrating once again yields 
that $\lim_{r \searrow \rho} \alpha =\pm \infty$, in contradiction to 
$\alpha(\rho)=0$.

Suppose now that $w_0=0$. We first consider the
$A$-equation~(\ref{eq:2.2}), which since $\omega=0$ becomes
\begin{equation}
r A' \;=\; 1-A \:-\: \frac{1}{e^2}\: \frac{(1-w^2)^2}{r^2} 
\:-\: \frac{2}{e^2}\:  A {w'}^2 \;.
    \label{eq:3.26}
\end{equation}
Employing the power ansatz (\ref{eq:Ap}),(\ref{eq:wp}) gives
\begin{equation}
    O(u^{s-1}) \;=\; 1 \:+\: O(u^s) \:-\: \frac{1}{e^2 r^2}\:+\: O(u^{2 \kappa}) \:+\: 
    O(u^{s+2\kappa-2}) \;. \label{eq:6b}
\end{equation}
Here and in what follows,
\[ f(u) \;=\; O(u^\nu) \spc{\mbox{means that}}\spc
\lim_{r \searrow \rho} u^{-\nu} \:f(u)\; {\mbox{ is finite and non-zero}}, \]
also we omit the expressions ``$o(u^\nu)$.'' The constant term in (\ref{eq:6b}) 
must vanish, and thus $e^2 \rho^2=1$. Using also that $O(u^s)$ is of higher 
order, (\ref{eq:6b}) reduces to
\begin{equation}
    O(u^{s-1}) \;=\; O(u) \:+\: O(u^{2 \kappa}) \:+\: 
    O(u^{s+2\kappa-2}) \;. \label{eq:63a}
\end{equation}
Suppose first that $s>2$. Then (\ref{eq:63a}) yields that $\kappa=\frac{1}{2}$. 
Substituting our power ansatz into the $Aw'$-equation (\ref{eq:2.5}) gives
\[ O(u^{s-\frac{3}{2}}) \;=\; O(u^{\frac{1}{2}}) \:+\: e^2 r\: T\: \alpha \beta \]
and thus $\alpha \beta \;=\; O(u^{\frac{1+s}{2}})$. Since $\beta(\rho) \neq 0$, 
we conclude that there are constants $c_1, \delta>0$ with
\begin{equation}
    |\alpha| \;\leq\; c_1 \: u^{\frac{1+s}{2}} \spc {\mbox{for 
    $\rho<r<\rho+\delta$}}.
    \label{eq:63b}
\end{equation}
From this one sees that the first summand on the right side of (\ref{eq:6a}) is 
of higher order; more precisely,
\[ \sqrt{A} \:\partial_r (\sqrt{A} \:\partial_r \alpha) \;=\;
O(u^{\frac{1}{2}}) \;. \]
Multiplying by $A^{-\frac{1}{2}}$ and integrating twice, we conclude that
\[ \alpha \;=\; O(u^{\frac{5}{2}-s}) \;, \]
and this contradicts (\ref{eq:63b}).

The final case to consider is $w_0=0$ and $s=2$. Now the $Aw'$-equation (\ref{eq:2.5})
gives
\[ O(u^\kappa) \;=\; O(u^\kappa) \:+\: e^2 T \:\alpha \beta \]
and thus $\alpha=o(u)$. This gives a contradiction in (\ref{eq:6a}) unless
$w-\sqrt{A}=o(u)$, and we conclude that $\kappa=1$.
Now consider the Dirac equation (\ref{eq:3.3}). Since $w(\rho)=0$, the 
eigenvalues of the matrix in (\ref{eq:3.3}) on the horizon are
$\lambda=\pm \sqrt{m^2+c^2/\rho^2}$. As a consequence, the fundamental solutions 
behave near the horizon $\sim u^{\pm \sqrt{m^2+c^2/\rho^2}}$. The boundary 
conditions (\ref{eq:5zz}) imply that $\alpha \sim u^{+\sqrt{m^2+c^2/\rho^2}}$, 
whereas $\beta, \gamma, \delta \sim u^{-\sqrt{m^2+c^2/\rho^2}}$, and we conclude 
that
\begin{equation}
    (\alpha \delta + \beta \gamma)(\rho) \;>\; 0 \;.
    \label{eq:63c}
\end{equation}
Next we consider the $AT^2$-equation (\ref{eq:2.6}), which for $\omega=0$ takes 
the form (\ref{eq:3.12}). It is convenient to introduce for the square bracket 
the short notation
\begin{equation}
[\:] \;=\; 2 m\:(\alpha^2-\beta^2+\gamma^2-\delta^2) 
+\frac{4  w}{r} \:\alpha \beta
+\frac{4c}{r}\:(\alpha \delta + \beta \gamma) \;.
    \label{eq:3.27a}
\end{equation}
We define the matrix $B$ by
\[ B \;=\; \left( \begin{array}{cccc}
    m & w/r & 0 & c/r \\ w/r & -m & c/r & 0 \\
    0 & c/r & m & 0 \\ c/r & 0 & 0 & -m \end{array} \right) \;. \]
A short calculation shows that
\[ [\:] \;=\; 2 \:\bra \Psi,\: B \:\Psi \ket \;, \]
and furthermore, using the Dirac equation (\ref{eq:3.3}),
\begin{eqnarray}
[\:]' & = & 2 \:\bra \Psi,\: B' \:\Psi \ket \;=\; 4 \alpha \beta \left(
\frac{w}{r} \right)^\prime \:-\:
\frac{4c}{r^2}\:(\alpha \delta + \beta \gamma) \nonumber \\
&=& \frac{4 w'}{r} \:\alpha \beta \:-\: \frac{4w}{r^2}\: \alpha \beta \:-\: 
\frac{4c}{r^2} \:(\alpha \delta + \beta \gamma) \;. \label{eq:bd}
\end{eqnarray}
Since $(\alpha \beta)(\rho)=0$ and $(\alpha \delta+\beta \gamma)(\rho) > 0$ 
according to (\ref{eq:63c}),
\[ - [\:]' \;\geq\; c_2 \spc {\mbox{for $\rho<r<\rho+\delta$}} \]
and a constant $c_2>0$ . Integrating on both sides shows that
\[ \left| [\:] \right| \;\geq\; c_3 \:u \spc {\mbox{for $\rho<r<\rho+\delta$}} \]
with $c_3>0$. As a consequence, the first summand in (\ref{eq:3.12}) diverges 
for $r \searrow \rho$, whereas the left side and the second summand on the 
right are bounded in this limit. This is a contradiction.

We conclude that $A^{-\frac{1}{2}}$ must be integrable near the event horizon, and so
$s<2$.

\section{Proof of the Main Theorem}
\label{sec7}
In this section we shall analyze the EDYM equations with the power ansatz
(\ref{eq:Ap}),(\ref{eq:wp}) near the event horizon. We will derive
restrictions for the powers $s$ and $k$ until only the exceptional
case~(\ref{eq:ex1}) of Theorem~\ref{thm:2.1} remains.
So far, we know from Section~\ref{sec6}
that $s<2$. A simple lower bound follows from the $A$-equation~(\ref{eq:2.2})
which for $\omega=0$ simplifies to (\ref{eq:3.26}). Namely in view of
hypothesis \AII, the right-hand side of (\ref{eq:3.26}) is bounded, and
thus $s \geq 1$. The case $s=1$ is excluded just as in~\cite{FSY2}
by matching the spinors across the horizon and applying a radial flux
argument. Thus it remains only to consider $s$ in the range
\begin{equation}
1 \;<\; s \;<\; 2 \;. \label{eq:3.27}
\end{equation}

We begin by deriving a power expansion for $\alpha$ near the event horizon.
\begin{Lemma}
\label{lemma71}
Suppose that $w_0 \neq 0$ or $\kappa \neq s/2$. Then
the function $\alpha$ behaves near the horizon as
\begin{equation}
\alpha \;=\; \alpha_0 \:u^\sigma \:+\: o(u^\sigma) \;,\spc
\alpha_0 \neq 0, \label{eq:7f}
\end{equation}
where the power $\sigma$ is either
\begin{equation}
\sigma \;=\; 1 - \frac{s}{2} \label{eq:7a}
\end{equation}
or
\begin{equation}
\sigma \;=\; \left\{ \begin{array}{cc} 2-s & {\mbox{if $w_0 \neq 0$}} \\
2-s+\min (\kappa, s/2) & {\mbox{if $w_0 =0$.}} \end{array} \right. \label{eq:7b}
\end{equation}
\end{Lemma}
{\Proof} We set
\begin{equation}
\sigma \;=\; \sup \left\{ p \: : \: \limsup_{r \searrow \rho}
|u^{-p} \:\alpha(r)| \:<\: \infty \right\} \;\leq\; \infty \;.
\label{eq:4m}
\end{equation}
Suppose first that $\sigma<\infty$.
Then for every $\nu<\sigma$ there are constants $c>0$ and $\varepsilon>0$
with
\begin{equation}
|\alpha(r)| \;<\; c \:u^\nu \spc {\mbox{for $\rho<r<
\rho+\varepsilon$}}. \label{eq:7d}
\end{equation}
We consider the first component of the squared Dirac equation
(\ref{eq:6a}) and write it in the form
\begin{equation}
\sqrt{A} \:\partial_r (\sqrt{A} \:\partial_r \alpha) \;=\; f\:\alpha+g
\;, \label{eq:7e}
\end{equation}
where $f$ stands for the square bracket and $g$ for the last summand in
(\ref{eq:6a}), respectively. Multiplying by $A^{-\frac{1}{2}}$ and
integrating gives
\[ \sqrt{A} \:\partial_r \alpha(r) \;=\; \int_\rho^r A^{-\frac{1}{2}} \:
(f \:\alpha+g) \:+\: C \]
with an integration constant $C$. We again multiply by $A^{-\frac{1}{2}}$
and integrate. Since $\alpha(\rho)=0$, we obtain
\begin{equation}
\alpha(r) \;=\; \int_\rho^r A^{-\frac{1}{2}}(s) \:ds \int_\rho^s
A^{-\frac{1}{2}} \:(f\:\alpha+g) \;+\; C \int_\rho^r A^{-\frac{1}{2}}
\;. \label{eq:77}
\end{equation}
Note that the function $f$, introduced as an abbreviation for the square
bracket in (\ref{eq:6a}), is bounded near the horizon. Hence (\ref{eq:7d})
yields a polynomial bound for $|f \alpha|$. Each multiplication with
$A^{-\frac{1}{2}}$ and integration increases the power by $1-\frac{s}{2}$,
and thus there is a constant $c_1$ with
\begin{equation}
\int_\rho^r A^{-\frac{1}{2}}(s) \:ds \int_\rho^s
A^{-\frac{1}{2}} \:|f\:\alpha| \;\leq\; c_1 \:u^{2-s+\nu}
\spc {\mbox{for $\rho<r<\rho+\varepsilon$}}. \label{eq:8a}
\end{equation}
Since $2-s>0$, (\ref{eq:8a}) is of the order $o(u^{1-\frac{s}{2}+\nu})$, and
thus (\ref{eq:77}) can be written as
\begin{equation}
\alpha(r) \;=\; \int_\rho^r A^{-\frac{1}{2}}(s) \:ds \int_\rho^s
A^{-\frac{1}{2}} \:g \;+\; C \int_\rho^r A^{-\frac{1}{2}} \:+\:
o(u^{1-\frac{s}{2}+\nu}) \;. \label{eq:7d2}
\end{equation}
Consider the behavior of the first two summands in (\ref{eq:7d2}).
The function $g$ stands for the last summand in (\ref{eq:6a}).
If $w_0 \neq 1$, it has a non-zero limit on the horizon. If on the other
hand $w_0=1$, then $g \sim u^\kappa$. Substituting into (\ref{eq:7d2}) and
integrating, one sees that the first summand in (\ref{eq:7d2}) is
$\sim u^\sigma$ with $\sigma$ given by (\ref{eq:7b}). The second summand
in (\ref{eq:7d2}) vanishes if $C=0$, and is $\sim u^\sigma$ with $\sigma$
as in (\ref{eq:7a}). According to (\ref{eq:3.27}),
$1-\frac{s}{2}<2-s<2-s+\min(\kappa, s/2)$.
Thus the values of $\sigma$ in (\ref{eq:7a}) and (\ref{eq:7b}) are different,
and so the first two summands in (\ref{eq:7d2}) cannot cancel each other.
If we choose $\nu$ so large that $1-\frac{s}{2}+\nu \geq \sigma$,
(\ref{eq:7d2}) yields the Lemma. 

Suppose now that $\sigma$ given by (\ref{eq:4m}) is infinite. Then choosing
\[ \nu \;=\; \max \left[ 1-\frac{s}{2}, \: 2-s+\min(\kappa, s/2) \right] \;, \]
we see that the first two summands in (\ref{eq:7d2}) are of the order
$O(u^s)$ with $s$ according to (\ref{eq:7a}) and (\ref{eq:7b}), respectively, 
and the last summand is of higher order. Thus (\ref{eq:7d2}) implies that 
$\sigma$ as defined by (\ref{eq:4m}) is finite (namely, equal to the minimum of
(\ref{eq:7a}) and (\ref{eq:7b})), giving a contradiction. Thus $\sigma$ is 
indeed finite.
\QED

In the proof of Proposition~\ref{prp53}, we already observed that the square
bracket in the $AT^2$-equation (\ref{eq:2.6}) vanishes on the horizon
(\ref{eq:4z}). Let us now analyze this square bracket in more detail, where we 
use again the notation (\ref{eq:3.27a}).

\begin{Prp} \label{prp38}
$\kappa<1$ and
\begin{equation}
[\:] \;=\; O(u^{\kappa+\sigma}) \:+\: O(u) \label{eq:3.33}
\end{equation}
with $\sigma$ as in Lemma~\ref{lemma71}.
\end{Prp}
{\Proof}
The derivative of the square bracket is again given by (\ref{eq:bd}). 
Now $\alpha_0 = 0$, $\beta_0 \neq 0$ and from Lemma~\ref{lem:3.4}, $\alpha_0 
\delta_0 + \beta_0 \gamma_0 \neq 0$; thus using (\ref{eq:Ap}),
(\ref{eq:wp}), and (\ref{eq:7f}), we get, for $r$ near $\rho$,
\begin{equation}
[\:]' \;=\; O(u^{\kappa-1+\sigma}) \:+\: u^{\sigma+(\kappa)} \:+\: O(1) 
\;,
    \label{eq:3.28}
\end{equation}
where we again omitted the expressions ``$o(u^.)$'' and we use the notation
\[ (\kappa) \;=\; \left\{ \begin{array}{ll}
\kappa & {\mbox{if $w_0=0$}} \\ 0 & {\mbox{if $w_0 \neq 0$}}
\end{array} \right. \;. \]
Integrating (\ref{eq:3.28}) and using that $[\:]_{r=\rho}=0$ according to
(\ref{eq:4z}), we obtain that
\begin{equation}
[\:] \;=\; O(u^{\kappa+\sigma}) \:+\: u^{\sigma+(\kappa)+1} \:+\: 
O(u) \;. \label{eq:3.30}
\end{equation}

Suppose $\kappa \geq 1$. Then $\kappa+\sigma>1$ and $\sigma+(\kappa)+1>1$,
and (\ref{eq:3.30}) becomes
\[ [\:] \;=\; O(u) \;. \]
We write the $AT^2$-equation (\ref{eq:3.12}) as
\begin{equation}
r\: (A T^2)' \;=\; T^3 \:[\:] \:-\: \frac{4 A \:T^2\:{w'}^2}{e^2}\;.
    \label{eq:3.32}
\end{equation}
Since $(A T^2)'$ is bounded and $T^3=O(u^{-\frac{3s}{2}})$
(by virtue of hypothesis \AI), (\ref{eq:3.32}) behaves near the event horizon
like
\begin{equation}
u^0 \;=\; O(u^{1-\frac{3s}{2}}) \:+\: O(u^{2\kappa-2})\;.
\label{eq:2r}
\end{equation}
Since $2\kappa-2 \geq 0$ and $1-\frac{3s}{2}<0$, the right side of
(\ref{eq:2r}) is unbounded as $r \searrow \rho$, giving a contradiction.
We conclude that $\kappa<1$.

For $\kappa<1$, the second summand in (\ref{eq:3.30}) is of higher order than 
the first, and we get (\ref{eq:3.33}).
\QED

In the remainder of this section, we shall substitute the power expansions
(\ref{eq:Ap})--(\ref{eq:Tp}) and (\ref{eq:7f}) into the EDYM equations and
evaluate the leading terms (i.e.\ the lowest powers in $u$). This will amount
to a rather lengthy consideration of several cases, each of which has
several subcases. We begin with the case {\underline{$w_0 \neq 0,\pm1$}}.
The $A$-equation~(\ref{eq:2.2}) simplifies to (\ref{eq:3.26}). The
$AT^2$-equation (\ref{eq:2.6}) for $\omega=0$ takes the form (\ref{eq:3.12}),
and we can for the square bracket use the expansion of
Proposition~\ref{prp38}. Finally, we also consider the $Aw'$-equation
(\ref{eq:2.5}). Using the regularity assumption \AI, we obtain
\begin{eqnarray}
{\mbox{$A$-eqn:}} && O(u^{s-1}) \;=\; 1 \:-\: \frac{(1-w_0^2)^2}{e^2 \rho^2}
\:+\: O(u^\kappa)
\:+\: O(u^{s+2\kappa-2}) \label{eq:Aa} \\
{\mbox{$AT^2$-eqn:}} && u^0 \;=\; O(u^{\kappa+\sigma-\frac{3s}{2}})
\:+\: O(u^{1-\frac{3s}{2}}) \:+\: O(u^{2 \kappa -2}) \label{eq:AT2a} \\
{\mbox{$Aw'$-eqn:}} && O(u^{s+\kappa-2}) \;=\; w_0\:(1-w_0^2) \:+\: O(u^\kappa)
\:+\: O(u^{\sigma-\frac{s}{2}}) \;. \label{eq:Awpa}
\end{eqnarray}
First consider (\ref{eq:Aa}). According to \AII, $s+2\kappa-2 \geq 0$, and
so all powers in (\ref{eq:Aa}) are positive. We distinguish between the cases
where the power $s+2\kappa-2$ is larger, smaller, or equal to the other powers
on the right of (\ref{eq:Aa}). Making sure in each case that the terms
of leading powers may cancel each other, we obtain the cases and conditions
\begin{eqnarray}
{\mbox{\bf{(a)}}} && \kappa < s+2\kappa-2
\;\;\Longrightarrow\;\; w_0^2=1 \pm e \rho,\;
\kappa=s-1, \;s \geq \frac{3}{2} \label{eq:7ca} \\
{\mbox{\bf{(b)}}} && \kappa = s+2\kappa-2
\;\;\Longrightarrow\;\; w_0^2=1 \pm e \rho,\;
\kappa=2-s, \;s \geq \frac{3}{2} \label{eq:7cb} \\
{\mbox{\bf{(c)}}} && \kappa > s+2\kappa-2 >0
\;\;\Longrightarrow\;\; w_0^2=1 \pm e \rho,\;
\kappa=\frac{1}{2}, \;s<\frac{3}{2} \label{eq:7cc} \\
{\mbox{\bf{(d)}}} && s+2\kappa-2 = 0
\;\;\Longrightarrow\;\; \kappa=1-\frac{s}{2} \;. \label{eq:7cd}
\end{eqnarray}
In Case~{\bf{(a)}}, the relations in~(\ref{eq:7ca}) imply that
\[ 1-\frac{3s}{2} \;<\; 2s-4 \;=\; 2\kappa-2 \;. \]
Hence (\ref{eq:AT2a}) yields $1-3s/2 = \kappa+\sigma-3s/2$, so
\begin{equation}
\sigma \;=\; 1-\kappa \;=\; 2-s \;.
    \label{eq:3.42}
\end{equation}
This is consistent with Lemma~\ref{lemma71}. But we get a contradiction
in (\ref{eq:Awpa}) as follows.
Since $\kappa=s-1$, we have $s+\kappa-2 = 2s-3 > 0$; on the other hand,
\[ \sigma-\frac{s}{2} \;=\; 2-s-\frac{s}{2} \;=\; 2-\frac{3s}{2} \;<\;0
\;. \]
Thus the left-hand side of (\ref{eq:Awpa}) is bounded, but the right-hand side
is unbounded as $r \searrow \rho$. This completes the proof in Case {\bf{(a)}}.

In Case~{\bf{(b)}}, (\ref{eq:Awpa}) yields that
\begin{equation}
\sigma \;\geq\; \frac{s}{2} \;. \label{eq:n2}
\end{equation}
We consider the two cases (\ref{eq:7a}) and (\ref{eq:7b}) in
Lemma~\ref{lemma71}. In the first case, (\ref{eq:n2})
yields that $s \leq 1$, contradicting (\ref{eq:3.27}). In the second case,
(\ref{eq:n2}) implies that $s \leq \frac{4}{3}$. This contradicts
the inequality in (\ref{eq:7cb}), and thus completes the proof in
Case {\bf{(b)}}.

In Case~{\bf{(c)}}, the relations in (\ref{eq:7cc}) give
$s+\kappa-2=s-\frac{3}{2}<0$, and thus~(\ref{eq:Awpa}) implies that
$s+\kappa-2=\sigma-\frac{s}{2}$, so $\sigma=\frac{3}{2}\:(s-1)$.
According to Lemma~\ref{lemma71}, $\sigma=2-s$ or $\sigma=1-\frac{s}{2}$.
In the first of these cases, we conclude that $s=\frac{7}{5}$ and
$\sigma=\frac{3}{5}$.
Substituting these powers into (\ref{eq:AT2a}), we get
\[ u^0 \;=\; O(u^{-1}) \:+\: O(u^{-\frac{11}{10}}) \:+\:
O(u^{-1}) \;, \]
which clearly yields a contradiction. Thus $\sigma=1-\frac{s}{2}$, giving
\begin{equation}
s \;=\; \frac{5}{4} \;,\spc \kappa \;=\; \frac{1}{2} \;,\spc
\sigma \;=\; \frac{3}{8} \;.
\end{equation}
This case is ruled out in Lemma~\ref{lemma73} below.

In Case~{\bf{(d)}}, we consider (\ref{eq:Awpa}). Since $s+\kappa-2=-\kappa<0$,
we obtain that $s+\kappa-2=\sigma-\frac{\sigma}{2}$ and thus
$\sigma=s-1$. Lemma~\ref{lemma71} yields the two cases
\begin{eqnarray}
s &=& \frac{4}{3} \;,\spc \kappa \;=\; \frac{1}{3} \;,\spc
\sigma \;=\; \frac{1}{3} \spc {\mbox{and}} \label{eq:7k} \\
s &=& \frac{3}{2} \;,\spc \kappa \;=\; \frac{1}{4} \;,\spc
\sigma \;=\; \frac{1}{2} \;. \label{eq:7l}
\end{eqnarray}
The first of these cases is the exceptional case of Theorem~\ref{thm:2.1},
and the second case is excluded in Lemma~\ref{lemma74} below.
This concludes the proof of Theorem~\ref{thm:2.1} in the case
$w_0 \neq 0,\pm 1$.

We next consider the case {\underline{$w_0=\pm1$}}. Then the expansions
(\ref{eq:Aa})--(\ref{eq:Awpa}) must be modified to
\begin{eqnarray}
{\mbox{$A$-eqn:}} && O(u^{s-1}) \;=\; 1 \:+\: O(u^{2 \kappa})
\:+\: O(u^{s+2\kappa-2}) \label{eq:Ab} \\
{\mbox{$AT^2$-eqn:}} && u^0 \;=\; O(u^{\kappa+\sigma-\frac{3s}{2}})
\:+\: O(u^{1-\frac{3s}{2}}) \:+\: O(u^{2 \kappa -2}) \label{eq:AT2b} \\
{\mbox{$Aw'$-eqn:}} && O(u^{s+\kappa-2}) \;=\; O(u^\kappa)
\:+\: O(u^{\sigma-\frac{s}{2}}) \;. \label{eq:Awpb}
\end{eqnarray}
One sees immediately that, in order to compensate the constant term in
(\ref{eq:Ab}), $s+2\kappa-2$ must be zero. Hence $s+\kappa-2=-\kappa<0$, and
(\ref{eq:Awpb}) yields that $s+\kappa-2=\sigma-\frac{s}{2}$ and thus
$\sigma=s-1$. Now consider Lemma~\ref{lemma71}.
In case (\ref{eq:7a}), we get the exceptional case of Theorem~\ref{thm:2.1},
whereas case (\ref{eq:7b}) yields that
\[ s \;=\; \frac{3}{2} \;,\spc \kappa \;=\; \frac{1}{4} \;,\spc
\sigma \;=\; \frac{1}{2} \;. \]
This case is ruled out in Lemma~\ref{lemma74} below, concluding
the proof of Theorem~\ref{thm:2.1} in the case $w_0 = \pm 1$.

The final case to consider is \underline{$w_0=0$}. In this case, the expansions
corresponding to (\ref{eq:Aa})--(\ref{eq:Awpa}) are
\begin{eqnarray}
{\mbox{$A$-eqn:}} && O(u^{s-1}) \;=\; 1 \:-\: \frac{1}{e^2 \rho^2}
\:+\: O(u^{2 \kappa})
\:+\: O(u^{s+2\kappa-2}) \label{eq:Ac} \\
{\mbox{$AT^2$-eqn:}} && u^0 \;=\; O(u^{\kappa+\sigma-\frac{3s}{2}})
\:+\: O(u^{1-\frac{3s}{2}}) \:+\: O(u^{2 \kappa -2}) \label{eq:AT2c} \\
{\mbox{$Aw'$-eqn:}} && O(u^{s+\kappa-2}) \;=\; O(u^\kappa)
\:+\: O(u^{\sigma-\frac{s}{2}}) \;. \label{eq:Awpc}
\end{eqnarray}
If $s+2\kappa-2=0$, we obtain exactly as in the case $w_0 \neq 0,\pm 1$
above that $\sigma=s-1$. It follows that $\kappa>\frac{s}{2}$, and 
Lemma~\ref{lemma71} yields either the exceptional case of Theorem~\ref{thm:2.1}, 
or $s=2$, contradicting (\ref{eq:3.27}). If on the other
hand $s+2\kappa-2>0$, we can in (\ref{eq:Ac}) use the inequality
$s+2\kappa-2<2 \kappa$ to conclude that $s-1=s+2\kappa-2$ and thus
$\kappa=\frac{1}{2}$. Now $\kappa<\frac{s}{2}$, and Lemma~\ref{lemma71} together 
with (\ref{eq:Awpc}) yields the two cases
\begin{eqnarray*}
s &=& \frac{5}{4} \;,\spc \kappa \;=\; \frac{1}{2} \;,\spc
\sigma \;=\; \frac{3}{8} \spc {\mbox{and}} \\
s &=& \frac{8}{5} \;,\spc \kappa \;=\; \frac{1}{2} \;,\spc
\sigma \;=\; \frac{9}{10}\;.
\end{eqnarray*}
The first case is ruled out in Lemma~\ref{lemma73} below, whereas the
second case leads to a contradiction in (\ref{eq:AT2c}). This concludes
the proof of Theorem~\ref{thm:2.1}, except for the special cases treated
in the following two lemmas.

\begin{Lemma}
\label{lemma73}
There is no solution of the EDYM equations satisfying the power
ansatz~(\ref{eq:Ap}), (\ref{eq:wp}), and (\ref{eq:7f}) with
\[ s \;=\; \frac{5}{4} \;,\spc \kappa \;=\; \frac{1}{2} \;,\spc
\sigma \;=\; \frac{3}{8} \;. \]
\end{Lemma}
{\Proof}
Suppose that there is a solution of the EDYM equations with
\begin{eqnarray*}
A(r) &=& A_0 \:u^{\frac{5}{4}} \:+\: o(u^{\frac{5}{4}}) \\
w(r) &=& w_1 \:u^{\frac{1}{2}} \:+\: o(u^{\frac{1}{2}})
\end{eqnarray*}
with parameters $A_0, w_1 \neq 0$. Consider the $A$-equation (\ref{eq:3.26}).
The left side is of the order $(r-\rho)^{\frac{1}{4}}$. Thus the constant
terms on the right side must cancel each other.
Then the right side is also of the order $u^{\frac{1}{4}}$. Comparing
the coefficients gives
\[ \frac{5}{4} \:\rho\:A_0 \;=\; -\frac{1}{2e^2} \:A_0\:w_1^2 \;. \]
This equation yields a contradiction because both sides have opposite sign.
\QED

\begin{Lemma}
\label{lemma74}
There is no solution of the EDYM equations satisfying the power
ansatz~(\ref{eq:Ap}), (\ref{eq:wp}), and (\ref{eq:7f}) with
\begin{equation}
s \;=\; \frac{3}{2} \;,\spc \kappa \;=\; \frac{1}{4} \;,\spc
\sigma \;=\; \frac{1}{2} \;. \label{eq:7pv}
\end{equation}
\end{Lemma}
{\Proof} According to (\ref{eq:Ap}), we can write the function $\sqrt{A}$ as
\begin{equation}
\sqrt{A} \;=\; u^\frac{3}{4} \:+\: f(u) \spc
{\mbox{with}}\spc f=o(u^\frac{3}{4}). \label{eq:7x}
\end{equation}
Employing the ansatz (\ref{eq:Ap}),(\ref{eq:wp}) into the A-equation
(\ref{eq:3.26}), one sees that
\begin{equation}
A {w'}^2 \;=\; u^0 \:+\: u^\frac{1}{4} \:+\: o(u^\frac{1}{4}) \;.
\label{eq:7y}
\end{equation}
We solve for $(w')^{-1}$ and substitute (\ref{eq:7x}) to obtain
\begin{eqnarray}
\frac{1}{w'} \;=\; u^\frac{3}{4} \:+\: u \:+\: c_1\:f \:+\: o(u)
\label{eq:7w}
\end{eqnarray}
with a real constant $c_1$. Now consider the $AT^2$-equation
(\ref{eq:3.12}), which we write again in the form (\ref{eq:3.32}) and
multiply by $A$,
\begin{equation}
r \:A \:(AT^2)' \;=\; (AT^2)^{\frac{3}{2}}
\:A^{-\frac{1}{2}}\: [\:] \:-\:
\frac{4}{e^2} \:(AT^2) \:(A {w'}^2) \;. \label{eq:48}
\end{equation}
As in the proof of Proposition~\ref{prp38}, a good expansion for
the square bracket is obtained by integrating its derivative. Namely, according
to (\ref{eq:bd}),
\[ [\:]' \;=\; \frac{4}{r} \:w'\:\alpha \beta \:+\: u^0 \:+\: o(u^0) \]
and thus
\[ A^{-\frac{1}{2}}(r) \:[\:] \;=\; \frac{4}{r} \:A^{-\frac{1}{2}} \:
\int_\rho^r (w' \:\alpha \beta)(s) \:ds \:+\: u^{\frac{1}{4}} \:+\: o(u^\frac{1}{4})
\;. \]
Substituting into (\ref{eq:48}) and using \AI\ and (\ref{eq:7y}), we obtain
\[ A^{-\frac{1}{2}}(r) \:\int_\rho^r w'\:\alpha \beta \;=\; u^0
\:+\: u^\frac{1}{4} \:+\: o(u^\frac{1}{4}) \;. \]
We multiply by $\sqrt{A}$, substitute (\ref{eq:7x}) and differentiate,
\[ w' \:\alpha \beta \;=\; u^{-\frac{1}{4}} \:+\: u^0 \:+\:
c_2\:f' \:+\: o(u^0) \;. \]
Multiplying by $1/w'$ and using (\ref{eq:7w}) gives the following expansion for
$\alpha \beta$,
\begin{equation}
\alpha \beta \;=\; u^\frac{1}{2} \:+\: u^\frac{3}{4} \:+\:
c_3 \:u^\frac{3}{4} \:f' \:+\: c_4 \:u^{-\frac{1}{4}} \:f \:+\:
o(u^\frac{3}{4})\;.
\label{eq:7v} 
\end{equation}
Next we multiply the $Aw'$-equation (\ref{eq:2.5}) by $\sqrt{A}$ and
write it as
\[ r^2 \:\sqrt{A} \:(\sqrt{A} \:(\sqrt{A} \:w'))' \;=\;
e^2 r \: (\sqrt{A} T) \: (\alpha \beta) \:+\: u^\frac{3}{4} \:+\:
o(u^\frac{3}{4}) \;. \]
We apply \AI\ and substitute (\ref{eq:7x}), (\ref{eq:7y}), and (\ref{eq:7v}). 
This gives an equation of the form (modulo higher order terms),
\[ u^{\frac{1}{2}} + u^{\frac{3}{4}} + u^{\frac{3}{4}}\:f' + u f'
+u^{-\frac{1}{4}}\:f \:+\: u^0 \;=\; u^{\frac{1}{2}} + u^{\frac{3}{4}} + 
u^{\frac{3}{4}} \:f' + u^{-\frac{1}{4}}\: f \;. \]
The constant term $\sim u^0$ must vanish since all the other terms tend to zero as 
$u \to 0$. Furthermore, the $u^{\frac{1}{2}}$ terms must cancel because 
all the other terms are $o(u^{\frac{1}{2}})$. We thus obtain
\[ u^{\frac{3}{4}} \:f' \:+\: u^{-\frac{1}{4}} \:f \;=\; u^{\frac{3}{4}} \:+\: 
u \:f' \:+\: f \;, \]
so that
\[ u\: f' \:+\: f \;=\; u \:+\: u^{\frac{5}{4}} \:f' \:+\: u^{\frac{1}{4}} \:f 
\;=\; u \:+\: o(u) \;, \]
and we find that $f$ satisfies an equation of the form
\[ d_1 \:u\:f' \:+\: d_2 \:f \;=\; d_3\: u \:+\: o(u) \;. \]
A straightforward but tedious calculation yields that the coefficients $d_1$
and $d_2$ both vanish, and that $d_3$ is non-zero. This is a contradiction.
\QED

\section{The Exceptional Case}
\label{sec8}
In this section, we consider the exceptional case
\[ s \;=\; \frac{4}{3} \;,\spc \kappa \;=\; \frac{1}{3} \;,\spc
\sigma \;=\; \frac{1}{3} \;. \]
By employing the power ansatz~(\ref{eq:Ap}), (\ref{eq:wp}), and (\ref{eq:7f}) into
the EDYM equations and comparing coefficients (using Mathematica), we find that
the solution near the event horizon is determined by the five free parameters
$(\beta_0, \gamma_0, m, c, \rho)$. The remaining parameters are given by
\begin{eqnarray*}
\delta_0 &=& \sqrt{\gamma_0^2-\beta_0^2 + \frac{2 c}{rm}\:\beta_0 \gamma_0} \\
w_0 &=& - \frac{c\:\delta_0}{\beta_0} \\
A_0 &=& \frac{9 \beta_0}{r}\: \sqrt{\frac{m^2 r^2 \:\beta_0^2 \:-\: 2cmr \:
\beta_0 \gamma_0 \:+\: c^2 \:\gamma_0^2}{r^2 - (1-w_0^2)^2}} \\
\alpha_0 &=& -3 \:\frac{mr \beta_0 - c \gamma_0}{r \sqrt{A_0}} \\
w_1 &=& \frac{9\:\alpha_0 \beta_0}{2r\:A_0}
\end{eqnarray*}
Expanding to higher order, we obtain after an arduous calculation two further
constraints on the free parameters, thus reducing the problem to one involving
only three parameters.  We investigated this three-parameter space numerically
starting at $r_0=\rho+\varepsilon$ and found strong evidence that no global
black hole solutions exist.  Indeed, either the power ansatz was inconsistent
near the event horizon (that is, for $r$ close to $r_0+\varepsilon$ the numerical solution
deviated from the power ansatz, and became singular as
$\varepsilon \searrow 0$), or else the solutions developed a singularity for finite
$r$.\\[1em]
{\em{Acknowledgments:}} JS would like to thank the Max Planck Institute for
Mathematics in the Sciences, Leipzig, and in particular Professor E.\ Zeidler,
for their hospitality and generous support. FF and JS are grateful to
the Harvard University Mathematics Department for support.

The research of JS and STY was supported in part by the
NSF, Grant No.\ DMS-G-9802370 and 33-585-7510-2-30.

\end{document}